\documentstyle[12pt]{article}

\newcommand{\be}{\begin{equation}}
\newcommand{\ee}{\end{equation}}
\newcommand{\ba}{\begin{eqnarray}}
\newcommand{\ea}{\end{eqnarray}}
\newcommand{\la}{\lambda}
\newcommand{\al}{\alpha}

\begin{document}
\hoffset=-.4truein\voffset=-0.5truein 
\setlength{\textheight}{8.5 in}
\setcounter{equation}{0}
\begin{center}
\vskip 0.6 in
{\large \bf  {  Duality and replicas for a unitary matrix model}}
\end{center}
\vskip .3 in
\begin{center}
 {\bf E. Br\'ezin$^{a)}$} 
{\it and} {\bf S. Hikami$^{b)}$}
\end{center}
\vskip 5mm
\begin{center}
{$^{a)}$ Laboratoire de Physique
Th\'eorique, Ecole Normale Sup\'erieure}\\ {24 rue Lhomond 75231, Paris
Cedex
05, France. e-mail: brezin@lpt.ens.fr{\footnote{\it
Unit\'e Mixte de Recherche 8549 du Centre National de la
Recherche Scientifique et de l'\'Ecole Normale Sup\'erieure.
} }}\\
{$^{b)}$ Department of Basic Sciences,
} {University of Tokyo,
Meguro-ku, Komaba, Tokyo 153, Japan. e-mail:hikami@dice.c.u-tokyo.ac.jp}\\
\end{center}     
\vskip 3mm         
\begin{center}
{\bf Abstract}
\end{center}
\vskip 3mm
   In a generalized Airy matrix model, a power $p$ replaces the cubic term of 
the Airy model introduced by Kontsevich. The parameter $p$ corresponds to 
Witten's spin index in the theory of intersection numbers of moduli space of 
curves. A continuation in $p$ down to $p = -2$ yields a well studied unitary matrix model.  
The application of duality 
and replica to the $p$-th Airy model provides, through this equivalence,  a 
generating function for both the weak and the strong coupling expansions of 
the unitary model. We thereby recover and extend further the results for 
these expansions. 

\section{Introduction}
\setcounter{equation}{0}
\renewcommand{\theequation}{1.\arabic{equation}}

The unitary matrix model with an external source is defined by the integral over the unitary group $U(N)$ 
\be\label{eq1}
Z = \int dU e^{{\rm tr}( U C^\dagger + U^\dagger C)}
\ee
where $U$ is a $N\times N$ unitary matrix, $U^\dagger U = 1$.   The external source matrix
$C$ is a given fixed complex matrix. Gross and Witten \cite {GW} have studied the case 
in which $C$ is a multiple of the identity, $C = N/g^2$, and showed that, in the large N limit, 
there is a third order transition at some critical $g_c$.
 Br\'ezin and Gross \cite{BG} generalized the model to include an arbitrary source matrix $C$ and 
again found that there is a phase transition governed by the parameter ${\rm tr} (C^{\dagger} C)^{-1/2}$. 
 This model has a weak coupling expansion , when the eigenvalues $\la_i$ of $C^{\dagger} C$  are large, 
which involves powers and products of parameters \be t_m = \sum_i \frac{1}{\la_i^{m-1/2}} = {\rm Tr} (C^{\dagger}  C)^{-(m-1/2)}. \ee
It has also a strong coupling expansion in powers of products of the parameters ${ \rm Tr} (C^{\dagger }C)^m$.
The strong coupling expansion is straightforward and the coefficients involving powers of 
the $\la^m$ have been tabulated long ago, up to $m=5$ \cite {Samuel}. 
The weak coupling expansion can be found by different methods ; 
Gross and Newman have used Virasoro equations \cite{Gross}  and  
we reproduce their analysis in section 4.  
We will follow here a completely different approach:
 we shall based the analysis on another matrix model, 
which will be shown to be equivalent to (\ref{eq1}), 
whose interest is to show 
that the coefficients of the expansion of $\log Z$ in powers of the $t_m$'s are related to topological invariants, namely  the intersection numbers of the moduli space of $p$-spin curves on a Riemann surface, in the singular limit $p\to -2$. The same model will also be used to recover and extend the results of the strong coupling expansion. 

Indeed we shall demonstrate a remarkable equivalence of the unitary model (\ref{eq1}),  
with a matrix model given by  an integral over  $N\times N$ Hermitian matrices 
\be\label{eq2}
Z _B= \int dB e^{{\rm tr} \frac{1}{B} + k {\rm tr log} B + {\rm tr} B \Lambda}
\ee
where $B$ is a $N\times N$ Hermitian matrix, $\Lambda = C C^\dagger$, provided one tunes properly 
the parameter $k$.  
This equivalence was already discussed  by Mironov et al. \cite{MMS,AMM} through Virasoro equations and
determinant formulations. 
We will show that one can  recover it  from an explicit determinantal expression 
in terms of  Bessel functions \cite{Brower}, and from a comparison between 
the expansions based on (\ref{eq2}) with that of (\ref{eq1}). 
However the equivalence  requires some proviso, 
as the reader, puzzled by the singularities due to vanishing eigenvalues of $B$, may guess.  
We are in fact dealing here with matrices which are unitarily equivalent to complex diagonal matrices.  
Therefore if we wrote $B= UXU^{\dagger}$, where $X$ is diagonal with complex entries,
 the integral over the unitary group would yield a Jacobian which is the modulus square of 
the Vandermonde determinant $J = \vert \Delta(X)\vert ^{2}= \prod_{i<j} \vert x_i-x_j\vert^2$ 
and the integration over the $x_i$s are over specified contours in the complex plane. 

Although the equivalence holds only if the coefficient $k $ goes to zero in the weak coupling limit and 
to $-N$ in the strong coupling limit,  we keep it arbitrary at this stage.
The weak coupling expansion for $k=0$ gives the intersection numbers of the moduli space, and for 
$k\ne 0$, the expansion is related to the intersection numbers of the discretized moduli space
\cite{Chekhov1}.
The expression of (\ref{eq2}) is the special case, $p=-2$ and $r=-1$,  of a ($p,r$) model
\be\label{eq3}
Z_{(p,r)} = \int dB {\rm exp}( -\frac{1}{p+1}{\rm tr}B^{p+1} + \frac{k}{r+1}{\rm tr} B^{r+1} + {\rm tr} B \Lambda )
\ee
which generalizes Kontsevich model \cite{Kontsevich} $ p=2, k=0$. The model that we want to study 
involves an analytic continuation from positive integer values of $p$ and $r$ to negative values. 

The reason that we find the model (\ref{eq2}) interesting, 
in addition of the topological invariants that it allows one to compute, 
is that it leads to completely different ways of calculating explicit expressions for the expansions.  
 In our previous work we have discussed   a powerful duality between expectation values of characteristic polynomials 
\cite{BH1,BH2,BH3,BH4}, namely the relation: 
\be\label{dual}
\int_{n\times n} dM \prod_{\alpha=1}^N {\rm det}( \lambda_\alpha - M ) 
e^{-\frac{1}{2}{\rm tr} M^2 + {\rm tr} M A}
= \int_{N\times N} dB \prod_{j=1}^n {\rm det} ( a_j - i B ) e^{-\frac{1}{2}{\rm tr} B^2 + {\rm tr} B \Lambda}
\ee
where $\Lambda = {\rm diag}(\lambda_1,...,\lambda_N)$ and $A= {\rm diag}(a_1,...,a_n)$. 
Both sides are integrals over Hermitian matrices (properly normalized). 
The left-hand side is an $N$ point function for $n\times n$ matrices in an external 
given matrix source $A$, whereas the r.h.s. is an $n$-point function for $N\times N$ matrices, 
in the matrix source $\Lambda$. The identity (\ref{dual}) generalizes the relation between
Hermitian and Kontsevich-Penner models at $A=0$ \cite{Chekhov,Kharchev}.
The strategy that we use consists of the following successive steps:
\begin{itemize}
\item one tunes the numbers $a_j$ in the r.h.s. in order to generate,  in an appropriate 
large $n$ scaling limit, the $(p,r)$ model (\ref{eq3}).
\item in the Gaussian model of the left-hand side we know explicit exact expressions 
for the correlators \cite{BH1,BH5}
\be\label{U}
U(s_1,...,s_l) = < {\rm tr} e^{s_1 M} {\rm tr} e^{s_2 M} \cdots {\rm tr} e^{s_l M} >.
\ee
in which the expectation value is meant with the normalised weight proportional to 
$e^{-\frac{1}{2}{\rm tr} M^2 + {\rm tr} M A}$.
\item the replica method, which consists of repeating the eigenvalue $\la_{\alpha}$,  $n_{\alpha}$ 
times and to let $n_{\al}$ go to zero.  This is used to relate the characteristic polynomials of the 
l.h.s. to the usual correlation functions as in 
\be \lim_{N\to 0}\frac{1}{N}  \frac{d}{d\la} ( {\rm det}(\la-M) )^N = {\rm tr} \frac{1}{\la -M} \ee
The resolvent is related by Fourier transform to ${\rm tr} e^{sM}$ which enters into (\ref{U}). 
We will explain later the meaning of taking an  $N=0$ limit here in spite of the fact that 
in the right-hand-side of (\ref{dual}) $N$ is fixed.

\end{itemize}

In this article, we use this strategy and tune  the external source $A$ 
to generate the($p,r$) model of (\ref{eq3}). 
 The limit $p=-2$ and $r=-1$, obtained through a  nontrivial
continuation from positive values of $p$ and $r$, is of particular interest.  
This rather involved construction is finally compared with the known expansions of the unitary
model $Z_U$ of (\ref{eq1})  \cite{Samuel,Gross,MMS}  both in the  strong   and in the weak coupling phases. 
Our results  provide   series expansion for intersection numbers with $l$ marked points and arbitrary genus. 
 The  consistency  with  known results  
 confirms that the duality plus replica method may be succesfully extended   to  negative
values of $p$ and $r$ in  the ($p,r$) model  (\ref{eq3}).

\section{Dual model and its $l$-point functions}
\setcounter{equation}{0}
\renewcommand{\theequation}{2.\arabic{equation}}
 The starting point is the duality relation (\ref{dual})  
 \be\label{2eq4}
\int_{n\times n} dM \prod_{\alpha=1}^N {\rm det}( \lambda_\alpha - M ) 
e^{-\frac{1}{2}{\rm tr} M^2 + {\rm tr} M A}
= \int_{N\times N} dB \prod_{j=1}^n {\rm det} ( a_j - i B ) e^{-\frac{1}{2}{\rm tr} B^2 + {\rm tr} B \Lambda}
\ee
where $\Lambda = {\rm diag}(\lambda_1,...,\lambda_N)$ and $A= {\rm diag}(a_1,...,a_n)$ ; 
the proof is simple and can be found in \cite{BH1}.
We now tune the $a_j$ in the r.h.s. to produce the model (\ref{eq3}) . We use the identity  
\be\label{2eq5}
\prod_{j=1}^n {\rm det}(1- \frac{i}{a_j} B) = 
{\rm exp} ( - \sum_{j=1}^n \sum_{m=0}^\infty \frac{i^m}{m a_j^m} {\rm tr} B^m )
\ee
and fix now the sums $\sum_{j=1}^n \frac{1}{a_j^m}$. Let us call $\rho$ the first one($m=1$) 
\be \label{rho} \rho = i\sum_j \frac {1}{a_j}, \ee
 which  may be absorbed by a shift of $\Lambda$. Taking next $\sum_{j=1}^n \frac{1}{a_j^2 }= 1$ 
the term $-\frac{1}{2}{\rm tr}B^2$ in the exponential is cancelled. Assume $p<r$; we can cancel 
all the terms other than  ${\rm tr} B^p$ and ${\rm tr} B^r$ by chosing the $a_j$ so that  
$\sum_{j=1}^n \frac{1}{a_j^m}=0$, for all $m<r$, except $m=2$ and $m=p$. 
Note that this requires in general complex $a_j$. In order to eliminate the terms with $m$ 
larger than $r$, we have to consider the following large $n$ scaling limit : 
\begin{itemize}
\item assume each $a_j$ is of order $n^{1/2}$ so that $\sum_j \frac{1}{a_j^2}= O(1)$ and in fact we chose them so that this last sum is exactly equal to one.
\item we take the $\la_{\al}$ near $\rho$ in a range in which  $ \la_{\al}- \rho \sim n^{-1/2 +2/r}$ 
\item we take the matrix $B$ of order $n^{1/2-2/r}$. Then one has $(\sum 1/a_j^r ){\rm tr} B^r$ of order one
\item in this scale $\sum 1/a_j^m {\rm tr} B^m \sim n^{2(1-m/r)}$ goes to zero for $m>r$ and one can drop all the powers higher than $q$.
\item however the term  $\sum 1/a_j^p {\rm tr} B^p \sim n^{2(1-p/r)} $ grows with $n$, since $p<r$ 
and we have to tune the $a_j$ so that  $\sum 1/a_j^p $ instead of 
being of order $n^{1-p/2}$ is of order $n^{1-p/2 -2(1-p/r)}$. 
\end{itemize}

In this scaling limit, with the $a_j$ chosen as indicated, the r.h.s.  gives the model  $Z_{(p,r)}$ of (\ref{eq3}).

The method consists of using  now the left hand side of (\ref{2eq4}). There the duality left us with the Gaussian model 
\be\label{2eq6}
Z_{dual} = \int_{n\times n} dM e^{-\frac{1}{2}{\rm tr} M^2 + {\rm tr} M A}
\ee
with the matrix $A$ fixed as just discussed above.
The reason for using this dual model is that  this we have for  (\ref{2eq6}), explicit expressions for the $l$-point correlation functions 
\be
U(s_1,...,s_l) = < {\rm tr} e^{s_1 M} {\rm tr} e^{s_2 M} \cdots {\rm tr} e^{s_l M} > \ee
where the average stand for
$$ <X> = \frac{1}{Z_{dual}} \int dM X(M) e^{-\frac{1}{2}{\rm tr} M^2 + {\rm tr} M A}$$
The result,  derived in \cite{BH1,BH2},  is 
\be U(s_1,...,s_l) = \int \prod_{i=1}^l \frac{du_i}{2 i \pi} \prod_{i=1}^l \prod_{j=1}^n ( 1 + \frac{s_l}{u_l - a_j})
{\rm det}\frac{1}{(u_{m_1} - u_{m_2} + s_{m_1})} e^{\sum u_l s_l +\frac{1}{2}\sum s_l^2}
\ee
 in which the contour integrals  circle around all the poles $a_i$. 

Let us first consider the one point function $U(s)$.
\ba\label{eq7}
U(s) &=& \frac{e^{\frac{s^2}{2}}}{s} \oint \frac{du}{2 i \pi} e^{s u} 
\prod_{j=1}^n \left( \frac{u + s - a_j}{
u - a_j}\right)\nonumber\\
&=& \frac{1}{s}\oint \frac{du}{2 i \pi} e^{\frac{1}{2}s^2 + s u + \sum {\rm log}(\frac{u + s -a_j}{u - a_j})}
\ea
We use now the $a_j$ tuned as discussed above. Expanding the logarithm, we have

$$\log{(\frac{a_j-u-s}{a_j-u})} = -s\sum \frac{1}{a_j} -\frac{1}{2} (s^2+2us) - \sum_{m=3}^\infty \frac{1}{m }\sum_{j=1}^n \frac{1}{a_j^m}((u+s)^m - u^m)$$
Using the previous tuning for the $a_j$ in the large-$n$ scaling limit 
we obtain
\be\label{eq8}
U(s) = \frac{e^{is\rho}}{s}\oint \frac{du}{2 i \pi}
e^{\frac{1}{p+1}((u+s)^{p+1}-u^{p+1}) + \frac{k}{r+1}((u+s)^{r+1}- u^{r+1})}
\ee

For $l$-point function $U(s_1,...,s_l)$, we obtain similarly.
\ba\label{eq9}
U(s_1,...,s_l) &=& \oint \prod_{j=1}^l \frac{du_j}{2 i \pi}
e^{\frac{1}{p+1}\sum_j((u_j+s_j)^{p+1} -u_j^{p+1}) + \frac{k}{r+1}\sum_j((u_j+s_j)^{r+1}-u_j^{r+1})}
\nonumber\\
&\times& {\rm det}\frac{1}{u_i-u_k+s_i}.
\ea
We have studied earlier the case in which  $k=0$ and $p > 0$ ; there the  $l$-point correlation functions $U(s_1,...,s_l)$ are the generating
function of the intersection numbers of the moduli space of curves with $p$-spins \cite{BH3,BH4}.
\vskip 2mm
\section{Weak coupling expansion}
\setcounter{equation}{0}
\renewcommand{\theequation}{3.\arabic{equation}}

We now study  the model in which one continues from integer positive $p$ and $r$ down to 
   $p=-2$ and $r=-1$.  We have in this case,
\be\label{3eq9}
U(s) = \frac{1}{s}\oint \frac{du}{2 i \pi}
e^{- \frac{1}{u+s} + \frac{1}{u}  + k {\rm log}(u+s) - k {\rm log} u}
\ee
In the original unitary model (\ref{eq1}), a simple model of one plaquette lattice QCD,  the matrix $C$ is inversely proportional to the gauge coupling constant. Therefore the eigenvalues $\la_{\al}$ of $C^{\dagger} C$ are large in weak coupling, small in strong coupling. The variable$s$ in $U(s)$ is a Laplace transform of $\la$; therefore the weak coupling limit corresponds to large $\la$, hence small $s$.  We first rescale $u \to s u$.
\be\label{3eq2}
U(s) = \oint \frac{du}{2 i \pi}e^{\frac{1}{s u(u+1)}} (\frac{u+1}{u})^k,
\ee
since  we continue later in $k$ down to negative values, the contour integral is really an integral over the discontinuity along the real $u$-axis. We chose a countour in the $u$-plane which goes parallel to the imaginary axis through the point $u= -1/2$ at which $1/u(u+1)$ is maximum in the real direction and change variable  $u = 1/2(- 1+i/x\sqrt {s})$ : 

\be\label{3eq10}
U(s) = -\frac{1}{4\pi \sqrt{s}}\int \frac{dx}{x^2}
e^{-\frac{4x^2}{1+sx^2}}(\frac{ 1-ix\sqrt{s}}{1+ix\sqrt{s}})^k \ee
This integral contains a singular term proportional to $1/\sqrt{s}$ which we discard, i.e. 
in order to obtain the terms of the expansion with positive powers of $s$,  we consider instead
\be\label{3eq100}
U_{+}(s) = -\frac{1}{4\pi \sqrt{s}}\int \frac{dx}{x^2}
[ e^{-\frac{4x^2}{1+sx^2}}(\frac{ 1-ix\sqrt{s}}{1+ix\sqrt{x}})^k - e^{-4x^2}]
\ee
For $s$  small, we can  expand the integrand in powers of $s$  and compute the successive coefficients as Gaussian integrals. The result for the positive powers of $s$ takes the form
 \ba\label{3eq50}
&&U_{+}(s) = - \frac{1}{2\sqrt{\pi s}}\biggl( - s \frac{1}{8}(4 k^2 -1) + s^2 \frac{1}{3! 2^7}
(4k^2-1)(4k^2-9)
\nonumber\\
&&- s^3 \frac{1}{5! 2^9} (4k^2-1)(4k^2-9)(4k^2-25) \nonumber\\
&&+ s^4 \frac{1}{21\cdot 2^{18}}(4k^2 -1)(4k^2-9)(4 k^2-25)(4 k^2-49)\nonumber\\
&&- s^5 \frac{1}{135\cdot 2^{22}}(4 k^2-1)(4 k^2-9)(4 k^2-25)(4 k^2-49)(4 k^2-81) +
O(s^6)\biggl)\nonumber\\
\ea

If we set $k=0$ in this result the $(p,r)$ model of (\ref{eq3}) reduces to the $p$-th Airy model, with here $p=-2$. Let us verify that the coefficients of $s^m$ in the above series (\ref{3eq50}) do coincide with the intersection numbers for $p=-2$ of the $p$-th generalized Kontsevich model \cite{BH4}. Indeed for this model
we had found 
\ba
&&U(s) = \frac{1}{N s^{1+\frac{1}{p}} \pi} [ \Gamma(1 +\frac{1}{p}) - \frac{p-1}{24} y \Gamma(1 - \frac{1}{p})
\nonumber\\
&&+ \frac{(p-1)(p-3)(1 + 2 p)}{5! 4^2 \cdot 3} y^2 \Gamma(1 - \frac{3}{p}) + \cdots]
\ea
where $y = s^{2 + \frac{2}{p}}$. For  $p=-2$, this does agree with the series (\ref{3eq50}) for $k=0$.

In the next section we shall compare these results with the weak coupling expansion derived both  for the model defined by (\ref{eq2}) and for the unitary model (\ref{eq1}) and see that they agree with (\ref{3eq50}) in the limit $k=0$.  The alleged correspondence of these two models is in fact true in the limit $k\to -N$, but in this calculation we are focusing on the terms of the weak coupling expansion which contain only one single trace. The flow of internal index has only one cycle and in the zero replica limit only those terms survive. 

\vskip 2mm
\section{Exact calculation, Virasoro constraints  and comparison}
\setcounter{equation}{0}
\renewcommand{\theequation}{4.\arabic{equation}}

Instead of using duality and replica we return to the partition function $Z_B$ defined in (\ref{eq2}). Standard techniques allow us to  express $Z_B$  as a determinant of modified Bessel functions for arbitrary $k$. We shall compare the weak coupling expansion deduced from this explicit expression, with that obtained above from $U(s)$ using   the dual model. 
Denoting  the eigenvalues of $B$ by $x_j$ ($j=1,...N$), we can integrate over the unitary group in the matrix integral (\ref{eq2}). Given the non commutation of the matrices $B$ and $\Lambda$ this requires the use of the HarishChandra (Itzykson-Zuber) formula : there is thus a division of the Jacobian measure $\vert \Delta(x)\vert^2$ by $\Delta(x)$ . Once the integration over the unitary group is done we have
\be\label{4eq1}
Z_B = \int \prod_{i=1}^N dx_i \frac{{\overline{ \Delta(x)}}}{\Delta(\lambda)} 
\prod_{i=1}^N x_i^k e^{\sum_i \frac{1}{x_i} + \sum x_i \lambda_i} .
\ee
in which the complex eigenvalues $x_i$ circle around the origin in the complex plane. 
Changing  variables $x_i  \to  \frac{1}{\sqrt{\lambda_i}} e^{i\theta_i}$, the integrals over these angles produce 
 modified Bessel functions $I_m(z)$, and $Z_B$ has an  explicit expression as a 
a determinant,
\be\label{4eq2}
Z_B = \frac{1}{(\prod \lambda_i)^{\frac{k+1}{2}}\Delta(\lambda)}
{\rm det} \left( \frac{1}{\lambda_i^{\frac{j-1}{2}}} I_{k+j}(2 \sqrt{\lambda_i}) \right)_{i,j}
\ee
In the weak coupling region of large $\lambda_j$, we use the asymptotic expansion of the
modified Bessel functions:
\ba\label{4eq4}
&&I_l(2 \sqrt{\lambda}) = \frac{e^{2 \sqrt{\lambda} }}{\sqrt{4 \pi \sqrt{\lambda}}}
( 1 - \frac{l^2-\frac{1}{4}}{4 \sqrt{\lambda}}+ \frac{(l^2- \frac{1}{4})(l^2 - \frac{9}{4})}{
2! (4 \sqrt{\lambda})^2}\nonumber\\
&&- \frac{(l^2 -\frac{1}{4})(l^2 - \frac{9}{4})(l^2 -\frac{25}{4})}{
3! (4 \sqrt{\lambda})^3} + \cdots ) .
\ea
Retaining only the  leading terms in (\ref{4eq4}) for large $\lambda_j$, gives the term of order zero 
\be\label{4eq3}
Z_0 = \prod_{i<j}^N\frac{1}{\sqrt{\lambda_i} + \sqrt{\lambda_j}} \prod_{i=1}^N \frac{1}{{\lambda_i}^{\frac{k}{2}}}
e^{\sum_{i=1}^N 2 \sqrt{\lambda_i}}
\ee
This is the nothing but the genus zero contribution to the free energy $F$ found in  \cite{BG} for the unitary model. 

Keeping N finite, we can calculate a few terms in the expansion in inverse powers of $\lambda$: 
\ba\label{4eq5}
&&Z_B = Z_0 [ 1 -  \frac{(2 k + 2 N)^2  -1}{16} \sum_{i=1}^N \frac{1}{\sqrt{\lambda_i}} \nonumber\\
&&+
\frac{((2 k+ 2 N)^2 -1)((2 k+ 2 N)^2 - 9)}{512}(\sum_{i=1}^N  \frac{1}{\sqrt{\lambda_i}})^2   \nonumber\\
&&  + \frac{((2 k + 2 N)^2 -1) ((2 k + 2 N)^2 -9)}{3! 4^6}\nonumber\\
&&\times \left(- 8 \sum_{i=1}^N 
\frac{1}{(\lambda_i)^{\frac{3}{2}}}
+ ((2 k + 2 N)^2- 17) (\sum_{i=1}^N \frac{1}{\sqrt{\lambda_i}})^3\right)\nonumber\\
&& + \cdots ]
\ea
Note that $N$ appears always in  the combination $( k +  N)$.
Therefore, setting  $k=-N$ , all the  factors $(2 k + 2 N)$ vanish 
and we find that 
 the weak coupling expansion of the partition function 
$Z_B$  (\ref{eq2}), up to this order ,  agrees with the weak coupling expansion obtained above in (\ref{3eq50}) with the dual model at $k=0$, if we look only at the single trace terms.  Furthermore it agrees with the known weak coupling expansion of  the unitary model \cite{Brower} $Z_U$ defined  by  (\ref{eq1}).  This is as a manifestation of  the equivalence of 
$Z_U$ in (\ref{eq1}) and $Z_B$ in (\ref{eq2}) at $k=-N$. 

This same equivalence at  $k=-N$ in (\ref{4eq5}) can be established through the weak coupling expansion
 based on  Virasoro equations. The Virasoro equations for the partition function $Z_U$ in (\ref{eq1}) have been derived
by Gross and Newman \cite{Gross}. We quote here  their results; the unitary model satisfies the obvious equation
\be\label{4eq6}
\frac{\partial^2 Z_U}{\partial C_{ab}\partial C_{bc}^\dagger} = \delta_{a c} Z_U
\ee 
but $Z_U$ is only function of the eigenvalues  $\lambda_i$  of $C C^\dagger$ ; one thus obtains\cite{BG}
\be\label{4eq7}
\frac{\partial^2 Z_U}{\partial \lambda_a^2} + \sum_{a\neq b}\frac{1}{\lambda_a - \lambda_b}
(\frac{\partial Z_U}{\partial \lambda_a}- \frac{\partial Z_U}{\partial \lambda_b})
= \frac{1}{\lambda_a}( Z_U - \sum_b \frac{\partial Z_U}{\partial \lambda_b})
\ee
With $Z_U = Z_0 Y$, the Virasoro constraints \cite{Gross} take the form
\be\label{4eq8}
L_n Y = - \partial_n Y \hskip 3mm n\ge 0
\ee
with  the $L_n$ given by 
\ba\label{4eq9}
&&n=0 : \hskip 5mm \sum_{k=0}^\infty (k + \frac{1}{2})\tilde t_k \partial_k Y + \frac{1}{16} = - \partial_0 Y,\nonumber\\
&&n\ge 1: \hskip 5mm \sum_{k=0}^\infty (k + \frac{1}{2}) \tilde t_k \partial_{k+n} Y + \frac{1}{4}
\sum_{k=1}^n \partial_{k-1}\partial_{n-k}Y = - \partial_n Y .
\ea
where
\be\label{4eq10}
\tilde t_k = - \frac{1}{2 k + 1}\sum_b \frac{1}{\lambda_b^{k + \frac{1}{2}}}
\ee

Following our previous notation for the $p$-spin curves, we use instead the parameters 
\be\label{4eq11}
t_m = \sum_i \frac{1}{\lambda_i^{m-\frac{1}{2}}}.
\ee
Thus $\tilde t_0$ corresponds to $- t_1$ in our notation. There is no $t_0$ in the expansion of $Z$
for the unitary matrix model.
 Using  the Virasoro constraints, we can sove for the case in which all the $t_n$, except $t_1$, vanish. The equation (\ref{4eq9}) allows one to determine  this  expansion:
\be\label{4eq12}
Y = 1 + \frac{1}{16}t_1 + \frac{9}{512} t_1^2 + \frac{9\times 17}{3! 4^6} t_1^3 + \cdots
\ee
 These terms correspond to the intersection numbers for one, two and three marked points.
If  we set $k=-N$ in our previous result (\ref{4eq5}), in terms of Bessel functions,  
we find that the expression  coincides indeed with $Y$ given by  (\ref{4eq12}). This shows within this expansion the coincidence of the two models. 
Note that the successive terms are independent of $N$.The weak coupling expansion for  general $k$ is given by (\ref{4eq5}). 
The equation (\ref{3eq50}) was missing this $k+N$, replaced by $k$ ; this is  because we were using the replica limit to focus on single trace terms. But the  true correspondence between the unitary and B-model would yield   $k+N$ instead of $k$ in (\ref{3eq50}), as one sees on (\ref{4eq5})  and we have to set $k=-N$ at the end to recover the unitary model.

\vskip 3mm
\section{Strong coupling expansion}
\setcounter{equation}{0}
\renewcommand{\theequation}{5.\arabic{equation}}
\vskip 2mm
We now turn to  the strong coupling expansion (small $\lambda_j$)  for the B-model. We can first  use the representation in terms of Bessel functions. 
Next we shall compare with the calculation based 
on  the $l$-point correlation function  $U(s_1,...,s_l)$ for large $s$ and verify that they agree as expected.
The ascending series for  the asymptotic expansion of the modified Bessel function is given by
\be\label{4eq12b}
I_\nu (z) = (\frac{1}{2}z)^\nu \sum_{m=0}^\infty \frac{(\frac{1}{4}z^2)^m}{m! \Gamma(\nu+m+1)}
\ee
Putting this expansion into the determinant of (\ref{4eq2}), we find for $k=-N$ for the unitary matrix model,
\ba\label{4eq13b}
Z &=& C ( 1+ \frac{1}{N} \sum_i \lambda_i + \frac{1}{2(N^2-1)} 
(\sum_i \lambda_i)^2 - \frac{1}{2 N (N^2-1)}
\sum_i \lambda_i^2\nonumber\\
&+& \frac{N^2-2}{6 N (N^2-1)(N^2-4)} (\sum_i \lambda_i)^3 - \frac{1}{2 (N^2-1)(N^2-4)}(\sum_i \lambda_i^2)
(\sum_i \lambda_i)\nonumber\\
&+& \frac{2}{3 N (N^2-1)(N^2-4)}\sum_i \lambda_i^3
+ O(\lambda^4))
\ea
where $C$ is a constant term. 

We now consider the strong coupling expansion of the B-model (\ref{eq2}) from the correlation functions $U(s_1,...,s_l)$ in its  dual representation. Strong coupling means  small $\lambda$'s. We are thus looking at the large $s$ expansion for the  Fourier  conjugate  variable $s\sim \frac{1}{\lambda}$.

Let us begin with the one point function $U(s)$, given by (\ref{3eq2}). The shift $u\to (u-1)/2$ 
 gives 
\be\label{5eq1}
U(s) = \frac{1}{2}\oint \frac{du}{2 i \pi} e^{\frac{4}{s (u^2 - 1)}} \left(\frac{u+1}{u-1}\right)^k.
\ee
As before there is an $s$-independent constant if we replace the exponential by one which does not affect the coefficients of the expansion, and we consider thus 
\be\label{5eq2}
U_+(s) = \frac{1}{2}\sum_1^{\infty} \frac{4^m}{m! s^m} \oint \frac{du}{2 i \pi} \frac{1}{(u^2-1)^m}\left(\frac{u+1}{u-1}\right)^k.
\ee
The contour integral is around the origin, but since the integrand falls off at infinity, we can replace the contour by an integral over the discontinuities accross the cuts which run from $ (1 , \infty)$ and $(-\infty ,  -1)$.  Then one finds
\be\label{5eq3} \oint \frac{du}{2 i \pi} \frac{1}{(u^2-1)^m}\left(\frac{u+1}{u-1}\right)^k = -\frac{2}{\pi} \sin{\pi k} \int_1^{\infty} dx \frac {(x+1)^{k-m}}{(x-1)^{k+m}} 
\ee
which vanishes as it should when $k$ is a positive or negative integer. The last integral is an Euler beta function:

\be\label{5eq31}
\int_1^{\infty} dx \frac {(x+1)^{k-m}}{(x-1)^{k+m}}  = 2^{1-2m} (2m-2)!  \frac { \Gamma(-k-m+1) }{\Gamma(-k+m)} .
\ee

We now  go to $k=-N$ as  discussed earlier, dropping obviously the vanishing factor $\sin{\pi k}$ . Note that the Gamma functions give
\be \label{5eq32} 
\frac { \Gamma(N-m+1) }{\Gamma(N+m)}   = \prod_{l=1}^{m-1} \frac {1}{ N^2-l^2}
\ee
which yields only even powers of $1/N$ in a $1/N$-expansion, as it should for  the initial unitary model.
Finally, up to an additional constant and to an overall factor we find
\be \label {5eq4}
U(s) = \sum_1^{\infty} \frac {(2m-2)!}{m! s^m} \frac{1}{  \prod_{l=1}^{m-1} ( N^2-l^2)} .
\ee
This expression agrees, as claimed earlier, with the single trace terms of the strong coupling expansion of $\log Z$ (\ref{4eq13b}) obtained from the determinant of Bessel functions. 

The result is an asymtotic expansion, as exhibited by the presence of    De Wit-'t Hooft
poles in $N$  \cite{DeWit} for any integer $N$, which imply that the $1/N$ expansion is certainly only asymptotic. 
Since the parameter $s$ is a conjugate  Fourier variable, the use of one single $s$ in $U(s)$, rather than several as in $U(s_1\cdots s_l)$  limits ourselves in the strong coupling (small $\la$) region  to terms of the form
\be\label{5eq11}
\frac{1}{s^m} = \sum_i \lambda_i^m
\ee
If we wanted to compare our results with the full result for the unitary integral we would need multiple $s_i$ to generate terms of the form $ \sum_i \lambda_i^{m_1}  \sum_i \lambda_i^{m_2}...$.  Therefore the expression (\ref{4eq10}) has to be compared with  the strong coupling expansion 
of  the unitary model in which we would keep  only the terms with one single trace ($\sum_i \lambda_i^m = {\rm tr} (C^{\dagger} C)^m$).

The strong coupling expansion of the unitary model has been studied long ago and we have tables for the expansion due to the work of Samuel \cite{Samuel} who studied the model 
\be
Z_U^s = \int dU {\rm exp}[\beta {\rm tr} A U + \beta {\rm tr} B U^{\dagger}]
\ee
where the $\beta$ is the coupling constant.
The expansion has the form  
\ba\label{samuel}
{\rm log} Z_U^s &=& N^2 \sum_{n=1}^\infty \frac{(\beta)^{2n}}{n!} \sum_{\alpha_1,...,\alpha_n}
N^{2n-2} C_{(\alpha_1,...,\alpha_n)}^c (N) \nonumber\\
&\times& (\frac{{\rm tr} AB}{N})^{\alpha_1} (\frac{{\rm tr} ABAB}{N})^{\alpha_2}
\cdots (\frac{{\rm tr}(AB)^n}{N})^{\alpha_n}
\ea
with the restriction $\alpha_1+2\alpha_2+\cdots+n\alpha_n = n$. (The superscript $c$ in $C_{(\al_1,\cdots \al_n)}^c$ refers to the connected part, the expansion of $\log Z_U$ rather than $Z_U$). The calculations of \cite{Samuel} have been performed up to order $\beta^{10}$.
Without in fact losing generality we can substitute $\beta A = C, \beta B = C^{\dagger}$  and use those results for the unitary model (\ref{eq1}).  
Let us first focus on the coefficients with one single trace, namely $\al_n =1$ and $\al_l = 0 $ for $l<n$.  We denote the coefficient of ${\rm Tr} (AB)^n$ simply by  $C_n$   and we find  in reference \cite{Samuel} 
\ba\label{5eq17a}
   && C_1 = 1, \hskip 3mm C_2= -\frac{1}{N^2-1}, \hskip 3mm C_3 = \frac{4}{(N^2-1)(N^2-4)}
\nonumber\\
&& C_4 = - \frac{30}{(N^2-1)(N^2-4)(N^2-9)},\nonumber\\
&& C_5 = \frac{336}{(N^2-1)(N^2-4)(N^2-9)(N^2-16)}
\ea
These numbers  agree completely  with the coefficients of $s^{-m}$ in (\ref{5eq4})  and the comparison shows that  at all orders  the result for (\ref{samuel}) would give for the single trace terms. 
\be C_m = (-1)^{m-1} \frac {(2m-1)!}{m!}  \frac{1}{  \prod_{l=1}^{m-1} ( N^2-l^2)} .\ee

If we want to use the same strategy to compare our results with the terms of the strong coupling expansion which involve product of several traces (or powers of one trace), we have to consider higher point functions. For instance the product of two traces requires now to consider the large $(s_1,s_2)$ expansion of  the two point function $U(s_1,s_2)$. We know this function explicitely: it is given by the double integral (\ref{eq9}). It has a  connected part $U(s_1,s_2)$ which is obtained by taking the  cycle of length two in the determinant: 

\ba\label{5eq20}
U(s_1,s_2) &=& \oint du_1 du_2 \frac{1}{(u_2-u_1+s_2)(u_2 - u_1 - s_1)}
e^{\frac{s_1}{u_1 (u_1+s_1)}+ \frac{s_2}{u_2(u_2+s_2)}}\nonumber\\
&&\times (\frac{u_2+s_2}{u_2})^k (\frac{u_1+s_1}{u_1})^k
\ea
We follow the same strategy : change variables $u_1\to s_1u_1$, $u_2\to s_2u_2$ expand the exponentials $e^{\frac{1}{su(u+1)}}$  in powers of $1/s$, but  we have also to deal with the pole terms 
$\frac{1}{ s_2 (1+u_2)-s_1u_1}$ and  $\frac{1}{ s_2 u_2-s_1(1+u_1)}$.  There we take a dissymetric limit in which $s_2\gg s_1$ and replace\be\label{5eqA}
\frac{1}{s_2(1+u_2)-s_1u_1} = \frac{1}{s_2}\sum_{m=0}^\infty \frac{u_1^m}{(1+u_2)^{m+1}}(\frac{s_1}{s_2})^m 
\ee
\be\label{5eqB}
\frac{1}{s_2u_2- s_1(1+u_1)}=\frac{1}{s_2} \sum_{m=0}^\infty \frac{(1+u_1)^m}{u_2^{m+1}}(\frac{s_1}{s_2})^m 
\ee
The result is a sum of factorized contour integrals in $u_1$ and $u_2$ similar to the one that we encountered in (\ref{5eq2}). These integrals may be deformed into integrals  of discontinuities over the real axis, and after division by the factors $\sin{\pi k}$ we can continue to $k=-N$. The integrals of the discontinuities are of the type
\be \int_0^{\infty} dx \frac {x^{N-b}}{ (1+x)^{N-a}} = B(N-b-1, b-a-1),\ee
with the Euler beta function $B(p,q) = \Gamma(p)\Gamma(q)/\Gamma(p+q)$. 
The coefficient of $1/s_1^{m_1} 1/s_2^{m-2} $ is then given as an explicit sum : 
\be \label{sum} \sum_{p_1,p_2} \frac{ B(N-m_1-p_2,p_1+p_2 +2m_1 +1) B(N-m_2 +p_1 +1, 2m_2 -p_1-p_2 -1)}{(m_1+p_1+p_2 +1)! (m_2-p_1-p_2 -1)!}.\ee
In the reference \cite{Samuel} we read the first coefficients $C_{1,m}$ of $ (\frac{{\rm tr} AB}{N})
(\frac{{\rm tr}(AB)^{m}}{N})$ namely
\ba C_{1,1} &&= 1/(N^2-1) \nonumber \\ C_{1,2}&&= -12/(N^2-1)(N^2-4)\nonumber \\
C_{1,3}&&=120/(N^2-1)(N^2-4)(N^2-9) \nonumber \\ C_{1,4}&&=-1680/(N^2-1)(N^2-4)(N^2-9)(N^2-16).\ea
We now compare  with the coefficient of the term $\frac{1}{s_2 s_1^{m-1}}$ in the strong coupling expansion of $U(s_1,s_2)$ (\ref{sum}). For instance the coefficient of $1/s_1s_2^2$ appears as a sum of three terms
\ba\label{5eq600}
&&\frac{N^2}{s_2^2 s_1} [ \frac{1}{N^2(N^2-1)(N+1)(N+2)} + \frac{1}{N^2 (N^2-1)(N-1)(N-2)} \nonumber\\
&&+ \frac{2}{N^2(N^2-1)^2}] = 
\frac{1}{s_2^2 s_1}\frac{4}{(N^2-1)(N^2-4)};
\ea
note that the sum if a function of $N^2$ as it should since the unitary integral (\ref{eq1}) has an expansion in even powers of $1/N$. 
Let us go further and quote from \cite{Samuel} the first two $C_{2,m}$   coefficients of  $ (\frac{{\rm tr} (AB)^2}{N}) (\frac{{\rm tr}(AB)^{m}}{N})$
\ba \label{c2} C_{2,2}&&=\frac{18 (3 N^2-7)}{(N^2-1)^2 (N^2-4)(N^2-9)}\nonumber  \\C_{2,3}&&= -480\frac{(3N^2-13)}{(N^2-1)^2(N^2-4)(N^2-9)(N^2-16)}.\ea
The sums (\ref{sum}) lead to a coefficient of $1/s_1^2s_2^2$ equal to
\ba
&&\frac{N^2}{s_1^2 s_2^2}(\frac{5}{N^2(N^2-1)(N^2-4)(N+1)(N+3)} \nonumber\\
&&+ \frac{5}{N^2(N^2-1)(N^2-4)(N-1)(N-3)}
+ \frac{8}{N^2(N^2-1)^2(N^2-4)}) \nonumber\\
&&= \frac{1}{s_1^2 s_2^2}\frac{6(3 N^2-7)}{(N^2-1)^2 (N^2-4)(N^2-9)}
\ea
and for $1/s_1^3 s_2^2$
\ba
&&\frac{N^2}{s_2^2 s_1^3}(\frac{42}{N^2(N^2-9)(N^2-4)(N^2-1) (N+1)(N+4)}\nonumber\\
&&+ \frac{42}{N^2(N^2-9)(N^2-4)(N^2-1)(N-1)(N-4)}\nonumber\\
&& + \frac{60}{N^2(N^2-1)^2 (N^2-4) (N^2-9)}\nonumber\\
&&= \frac{1}{s_2^2 s_1^3}\frac{48(3 N^2-13)}{ (N^2-1)^2 (N^2-4)(N^2-9)(N^2-16)}.
\ea
Again the various sterms of the sums combine and give, as expected, functions of $N^2$. The comparison with Samuel's results (\ref{c2}) is remarkable. Still there are combinatorial factors which make the agreement incomplete : for instance a factor 10 is missing in the expansion of $U(s_1,s_2)$ for $C_{2,3}$  which corresponds to the choice $5!/2!3!$ of assigning $s_1$ and $s_2$ to a particular trace, i.e. the number of ways of splitting 5 objects into two groups of of 2 and 3 ; a factor 3 has to multiply our coefficient of $1/s_1^2s_2^2$, i.e. the number of ways of splitting the 4 powers in 2 and 2. More generally a factor $(m_1+m_2)!/m_1!m_2!$ if $m_1\neq m_2$ and $ (2m_1)!/2(m_1!)^2$ if $m_1=m_2$.

Clearly this strategy may be extended  to $l$-point functions ($\ge3$),  for any $l$.  For instance,  if $l=3$, we take the connected part of the 3 by 3 determinant in (\ref{eq9}),
\ba\label{5eq100}
&&\frac{1}{(u_2-u_1+s_2)(u_3-u_2+s_3)(u_1-u_3+s_1)} \nonumber\\
&& + \frac{1}{(u_3-u_1+s_3)(u_2-u_3+s_2)(u_1-u_2+s_1)}
\ea
and repeat exactly the same expansions with some ordering of $s_1, s_2, s_3$ as above, to obtain finally the coefficient 
of $s_1^{-m_1}s_2^{-m_2}s_3^{-m_3}$.  For instance, we obtain 
at the lowest order  the coefficient of $s_1^{-1} s_2^{-1} s_3^{-1}$ as
\be
-\frac{1}{s_1 s_2 s_3} \frac{8}{ (N^2-1)(N^2-4)}
\ee
which, up to a sign, coincides with  $C_{1,1,1}$, the coefficient of $({\rm Tr} (AB))^3$. 

We have thus verified  that  the functions $U(s_1,...,s_l)$ provide the generating function for the coefficients $C_\alpha^c(N)$ for finite $N$. 
The strong coupling expansion,  in the presence of the external source $C^{\dagger}C$ has  also been analysed by Virasoro type 
recursion relations \cite{MMS}. However our explicit formulae for $U(s_1,...,s_l)$, from which one may obtain both the strong coupling and the weak coupling expansions, are very powerful. 

\section{Summary}
\setcounter{equation}{0}
\renewcommand{\theequation}{7.\arabic{equation}}
\vskip 2mm

We have found that the unitary matrix model may be replaced by a generalized $p$-spin higher 
Airy model which yields the same weak and strong coupling expansions. After a duality which replaces 
this alternative $(p,r)$ model by a Gaussian model in an external matrix source, 
we can compute explicitely the Fourier transform of the $l$-point correlation functions $U(s_1,...,s_l)$ 
(\ref{eq9}). This provides 
when $p=-2$ and $r=-1$  a generating function for the
  weak and the strong coupling expansions of the unitary matrix model $Z_U$ in (\ref{eq1}). The number $l$
correponds to the total number of traces  involving the external source matrices $C C^{\dagger}$ in 
the expansions.

In the weak coupling region, the replica method provides the terms of the expansion ordered by the number of independent traces involved in the expansion. The zero-replica limit for the one-point function $U(s)$, determines the single trace terms and we have verified the agreement with the weak coupling expansion of the unitary matrix model.

In the strong coupling region, the $l$-point functions $U(s_1,...,s_l)$  remarkably produce the correct
results of the strong coupling expansion of the unitary matrix model with an external source, 
with its involved $N$-dependence, De Wit-'t Hooft poles,  
and allows one to go much beyond the terms which had been tabulated long ago. 

Therefore this other matrix model  provides finite integrals, on one variable for 
the single trace terms, two variables for product of two traces, etc.,   which yield   
 both the weak and the strong coupling regions.  
It also shows an unexpected connexion between the unitary matrix model 
and the generalized Kontsevich model which we know is a generating function for intersection numbers
\cite{Kontsevich,Witten1,Penner,HZ}. 
 
The Gross-Witten phase transition  between the two regions  may also be 
recovered on the integrals for $U(s)$ ; it comes from the logarithmic potential of 
the matrix model (\ref{eq2}). The applications  of these techniques to other values 
of $p$ than $-2$ with a logarithmic 
potential may be interesting, and  it leads also to
integral representations for $U(s_1,...,s_l)$ as in (\ref{eq9}).

  Finaly let us note the phase transition in the unitary matrix model has been discussed over 
the last years by several authors in the context of string/black hole physics 
\cite{Witten2,Susskind,Wadia}. 
The reformulation presented here may be of interest in this other context as well.  

\vskip 3mm
{\bf Acknowledgement} S.H. is  supported by Grant-in Aid for Scientific Research(C) of JSPS.


\end{document}